\pdfoutput=1 %arXiv admin added this line
\documentclass[12pt]{article}
\usepackage{mathtext}
\usepackage{graphicx}
\usepackage{amsfonts}
\usepackage{amssymb}
\usepackage{amsmath}
\newcommand{\nn}{\medskip}

\newcommand{\ar}{\longrightarrow}

\begin{document}
\title{Quantum selfish gene (biological evolution in terms of quantum mechanics)}

\author{Yuri I. Ozhigov\thanks{ozhigov(at)cs.msu.su}, \\
Moscow State University of M.V.Lomonosov, Faculty VMK, \\chair of supercomputers and quantum informatics \\
}
\maketitle
PACS: 03.65,  87.10\\

\begin{abstract}
I propose to treat the biological evolution of genoms by means of quantum mechanical tools. We start with the concept of meta- gene, which specifies the ''selfish gene'' of R.Dawkins. Meta- gene encodes the abstract living unity, which can live relatively independently of the others, and can contain a few real creatures. Each population of living creatures we treat as the wave function on meta- genes, which module squared is the total number of creatures with the given meta-gene, and the phase is the sum of ''aspirations'' to change the classical states of meta- genes.  Each individual life thus becomes one of possible outcomes of the virtual quantum measurement of this function. The evolution of genomes is described by the unitary operator in the space of psi-functions or by Kossovsky-Lindblad equation in the case of open biosystems. This operator contains all the information about specific conditions under which individuals are, and how ''aspirations'' of their meta- genes may be implemented at the biochemical level. We show the example of quantum description of the population with two parts of meta-gene: ''wolves'' and ''deer'', which can be simultaneously in the same abstract living unity. ''Selfish gene'' reconciled with the notion of individuality of alive beings that gives possibility to consider evolutionary scenarios and their possible physical causes from the single position.
\end{abstract}

\section{Why quantum approach?}

Yet Schrodinger noted the quantum nature of life in his book
''What is life'' (\cite{Shr}), and with the development of biochemistry it became apparent. Quantum mechanics is not just a physical theory. This is the only theory that describes all the phenomena of electromagnetism, including chemistry and biology
at the level of accuracy that is now generally available, and its further expansion is restrained only by the complexity of computational problems. In this article I will try to show how biological evolution (\cite{We}, \cite{Fi}) may be expressed in its language. 

Here, very briefly, the essence of the quantum approach. Real particle (for example, an electron) is represented not as a point particle, but as a swarm of point particles called its samples. The object that we call a real particle, there is some semblance of a swarm of bees, in which each bee - a full-fledged representative of a swarm. In the experiments, we always ''see'' an electron at one point, it never ''splits'', as well, and one bee can not live without the swarm.

Every single sample has its own trajectory, while driving along which it accumulates a certain amount, called the amplitude $\Psi$, just as the bee collects nectar, where the initial amplitude is given at the starting point at the initial time. 

In the final time at any point in space occurs the so-called interference, which looks as follows. The amplitudes of all samples who find themselves at this point are added. The resulting number is called the resultant amplitude. The real particle is at the same time in different points of space, but at each point with the corresponding amplitude, which is calculated according to the law of interference. The amplitude is a complex number such that the square of its modulus is the probability density to find our real particle at a given point. The density of quantum swarm thus equals $|\Psi |^2$.

This is the Great Quantum Law of Nature, from which deviations - even the slightest - not detected! 

What is important is that the drops of ''nectar''
which was collected by our imaginary bees are complex numbers. Therefore, when adding the contributions of different ''bees'' at any point there is a whole set of different opportunities from fully
constructive interference (when all the vectors are similarly directed) to completely destructive (when their directions are different, and their sum is the small number). But this feature is not the most important.
The fact is that the function $ \Psi $ of the closed quantum system obeys Schrodinger equation
\begin{equation}
i\dot\Psi=H\Psi ,
\label{Sh}
\end{equation}
where the operator $ H $ in the simplest case of a single particle is
proportional to the second derivative of the function $ \Psi (x, t) $ with respect to the spatial coordinate $ x $. This equation - the only
 short equation which can describe the {\it evolution}
 rather than a simple dynamics.
To understand how the {\it evolution}  differs from the simple dynamics
compare Schrodinger equation (\ref{Sh}) with similar real
equations of heat conductivity (\ref{He}, 1)) and oscillations (\ref{He},
2)) where $ H $ is the same as in Schrodinger equation, but the left side has no imaginary unit:
\begin{equation}
1)\ \dot\Psi=H\Psi ,\ \ \ 
\label{He}
2) \ \ddot\Psi=H\Psi ,
\end{equation}

The last two equations describe a simple dynamics: first -
statistical (process of heat transfer), the second - ordinary
(oscillation of an elastic rod). We can always assume that the coordinate $ x $ is
some kind of material that is subject to change with
time, whereas the unknown function  in any equation is the quantity of this material.  The last two equations have an ensemble representations as well, where $x$ plays the role of coordinates of their samples. The essence of simple dynamics is that only the value of the unknown function itself has the speed and momentum, whereas in the quantum swarm speed and inertia is a property of the samples.
This is why in the case of simple dynamics reducing the grain of spatial resolution always leads to the required increase in
 accuracy.  In biology, the genome can act as $ x $, the value of the material (more precisely $ | \Psi |^2 $) is the number of individuals with a given genome.

In the equations (\ref{He}) the change of unknown function is becoming more and more smooth with decreasing grain of spatial resolution. We can say that the dynamics is the smooth change of $\Psi $ along the y-axis.
If $ x $ is a member of any team, and $\Psi $ is his
reward (or punishment) for the quality of work, simple
dynamics is a policy of penalties and rewards. This policy does not lead to any {\it  evolution}. (Mathematically, I could express it as the relative stability of direct difference scheme for the equations (\ref{He}, 1), 2)), if it is for someone to be clearer).

Now let us turn to Schrodinger equation (\ref{Sh}). Here character of change of $ \Psi $ with time will be quite different. To understand this, we turn to Bohm quantum hydrodynamics (\cite{Bo}), which is the equivalent form of Schrodinger equation. Its essence is that quantum swarm can be considered as an ensemble of ordinary point particles moving in the additional quantum pseudo-potential, which is proportional to 

\begin{equation}
-|\Psi |''_{xx}/|\Psi |.
\label{hydro}
\end{equation}

 This means that the quantum swarm is under the influence of special force outgoing from the swarm itself and not from the external potential. 
This force can be infinite, and shifts all samples found in any small cube with the side $dx$ towards increasing of the second derivative of psi-function.

This has serious implications. In the graph $ \Psi (x, t) $ will happen the superimpositions of one vertical layer to another, so that there will be ''peaks''. In each of these peaks, the second derivative will be infinite, that at the next step will lead to their ''explosion''. (Explosive nature of quantum evolution is a manifestation of the Bohr- Heisenberg uncertainty relation, see, e.g., (\cite{Fe}). It does, for example, incorrect the use of direct difference scheme to the Schrodinger equation - it simply diverges). Swarm shift along the x-axis - this is the real {\it evolution}. It
can not be obtained by ''rewards and punishments''. Going on here something qualitatively different: members of the team are simply dismissed and are recruited again, but in other places. Adequate mathematical apparatus for the description of biological evolution can thus be obtained, starting from quantum theory.

\begin{figure}
\includegraphics[scale=0.4, 
bb=5 10 483 600]{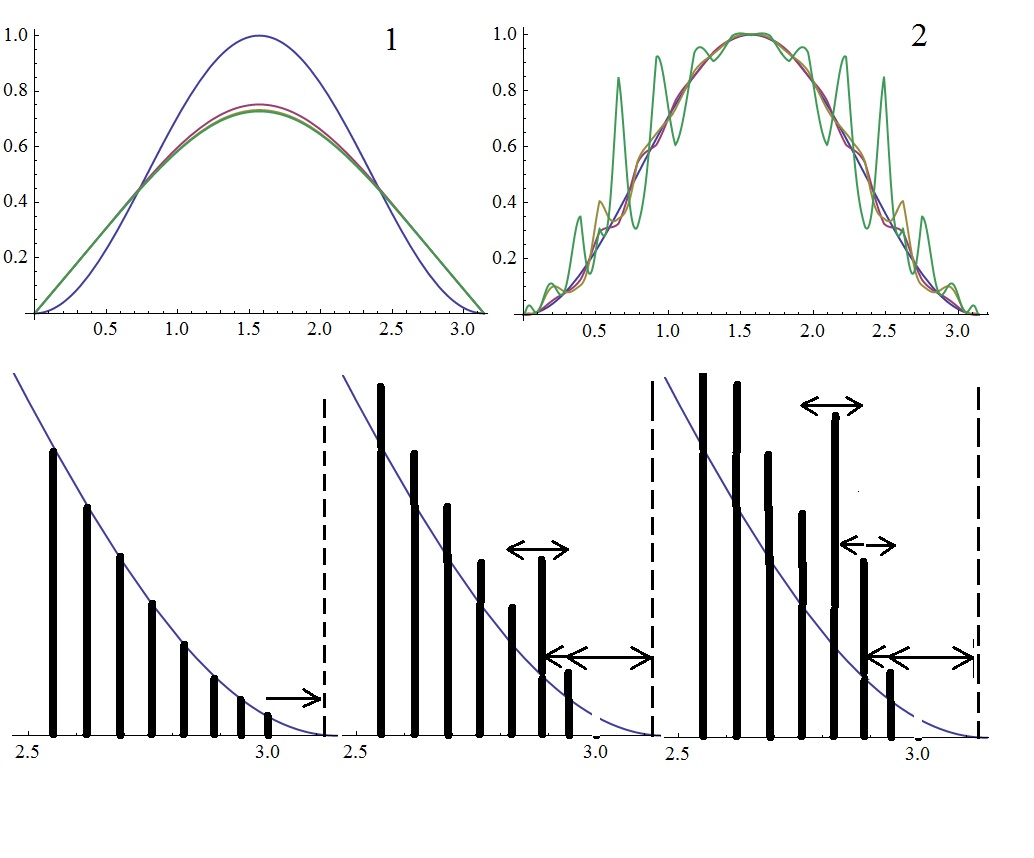}
 \caption{{\bf Above}: comparison of solutions of heat conduction equation (1) and Schrodinger equation (2), obtained by direct difference scheme with the same initial and boundary conditions. 1. The process of cooling of the rod, a smooth change of the ordinate, finite difference scheme for a long time does not decay. 2. - {\it Evolution} of psi-function of a particle in an infinitely deep potential well. Direct difference scheme quickly
 decays due to the rapid emergence of'' peaks''. Heat and oscillation equations represent the simple dynamics, and
Schrodinger equation represents the {\it evolution}. {\bf Below}: The mechanism of the appearance of ''peaks'' in the process of {\it evolution}. Movements of the vertical sections of the swarm along the x-axis generates'' peaks''.'' Peaks'' explosions launch a chain reaction of chaotic behavior
of the quantum swarm.}
\end{figure}

\begin{figure}
\includegraphics[scale=0.4, 
bb=5 10 383 530]{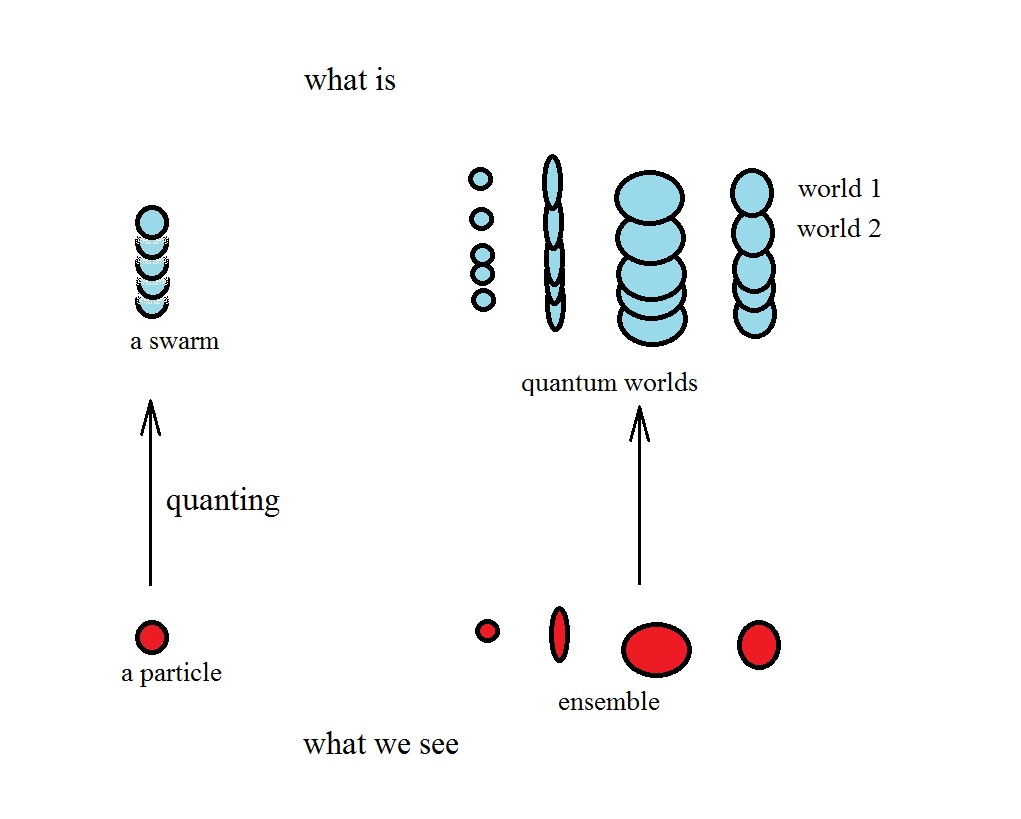}
 \caption{Quantization in terms of the swarm. Below - the classical states, above - the quantum states of the same objects.}
\end{figure}

The Great Law is generalized to systems of many particles, only here the role of ''particle'' will play an ensemble of $ n $ particles. Samples of such ''particle'' will be (virtual) ensembles, each of which occupies a special place in space, so that the phrase ''to be at a given point'' means the pairwise coincidence of coordinates of all corresponding individual particles of two such ensembles.

 The probability of such a ''collision of quantum worlds'' for arbitrary complex systems falls exponentially with increasing the system, so there is a very sharp barrier between the types of systems in which a collision plays a creative role (and therefore, it makes sense to apply quantum mechanics) and the other in which dominates quantum chaos (\cite{La}).

The first type includes systems, allowing the so-called canonical transformation, encapsulating the quantum complexity (entanglement) within the quasi-particles, for example, acoustic oscillations in a solid or quanta of the electromagnetic field - photons. There are reasons to believe that living matter belongs to this type of systems. Entanglement (see \cite{As},\cite{Ze}) is the single known source of negative relative entropy, which can create the order from the chaos that is the key property of living matter. 

Barrier of creativity is very serious, it delineates an area in which it makes sense to spend a lot of force on quantum mechanics at all.
Crossing over this barrier  causes difficulties in the building of quantum computer by type of microelectronics. Attempts to immediately get to the interference of quantum worlds in biology (see, eg, \cite {Og}, where they tried to consider the superposition of normal and mutant DNA) have not been successful precisely because underestimation of the sharpness of the creativity barrier. Analogue of the canonical transformation for the living matter is not so simple, and its search should be based on a purely biological foundation. The quantum nature of biology lies deeper than penetration zone of traditional physical instruments, the mathematical theory is thus more important than the direct physical experiments on living.

"Quantum computer works in the head of any of us, we just have to understand - how!" (Kamil Valiev). And this quantum computer - not
 microelectronic, not of Feynman version. It is biological.
Its main purpose - not solving math problems like search (\cite{Gr}, \cite{Sh}). Its destination - to manage living
substance and its biological evolution.

\section{Space of quantum states of living}

Genes will play the role of classical states of real particles, and organisms (or rather, generalized organisms filled with meta-genes) - the  role of their
samples (see. \cite{Wi}, and also \cite{Ha},  \cite{Do} - gene is the main participant of biological evolution, while living beings are only instruments serving him for reproduction; see also criticism \cite{Ma},\cite{Go},\cite{El}).

Coordinate $ x $ of psi-function must include not
only the spatial position of the living creatures, but also the state of
their genomes.

It may be reasonable to extend the concept of the genome by including also some variable part, for example, the quantum state of the DNA molecule, which is described in the language of quasi-particles (vibrational motion of the nuclei of atoms, electrons and their spins, see \cite{Win}). Language of quasi-particles allows to include electromagnetic field to the model and potentially take into account even cosmic factors of the evolution (see, for example, \cite{Shn}).
Quasiparticles exist in polymer molecules where rigid
bonds between adjacent monomers in a linear chain decrease entropy of
systems of such molecules so that the small energy pulses can greatly affect the dynamic scenario (see \cite{Kho}). The variable part of the genome in the form of quasi-particles can serve as an internal physical mechanism of directed DNA changes leading to biological evolution, along with an external factor - viral and bacterial environment. Hidden nature of internal factors can lead to apparently sharp genetic jump, which gives a new quality of organisms in the apparent absence of intermediate forms of living beings. 

{\bf Meta-gene is a classical
state of hereditary material of an organism, which is sufficient for its
independent living.} This includes DNA, as well as possibly the mitochondrial DNA, and possibly the DNA or RNA of viruses or some bacteria inhabiting organisms.

Here a ''body'' does not necessarily mean one
 individual, for example, for creatures with sexual reproduction is permissible
the distribution of meta-gene between male and female individuals. And vice versa: few meta-genes can exist in one real body, like in the following example where ''wolves'' and
''deer'' should be interpreted that way; they both can
belong to the same body (see picture 3). 

\begin{figure}
\includegraphics[scale=0.4, 
bb=5 10 383 530]{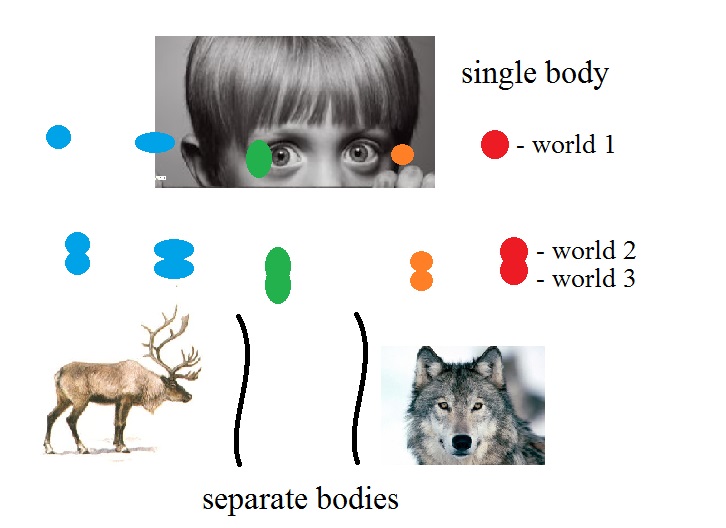}
 \caption{Meta-gene consists of different parts, which can belong to one or several bodies.}
\end{figure}

We select two intervals: $ \tau $ - the time of meta-gene expression, and $ T $ -
duration of a step of biological evolution, so that the second much
superior to the first, and the first is comparable to the lifetime of
separate individual. 

Virtual states provide a huge variety of genes instead of
strings of letters $A,G,T,C$ used by geneticists. True selection, which determines the evolution takes place in a tank of these states. But it is hidden
from traditional measurements exactly the same as the true state of
a quantum computer is hidden from the observer. 

We fix a set of meta-genes $ g_1, g_2, \ldots, g_n $, each of which is stable over many generations. We shall treat each meta-gene
as the self-sufficient model of the universe, so that the interaction between
different meta-genes will be an exact analogue of interference between quantum worlds (see picture 2). In this case the actual microscopic content of the evolution can be localized precisely within individual, where the worlds collide, leading to interference of amplitudes. 

Separate individual capable of independent life, typically has more than one meta-gene, but a few. The individual is the battlefield of meta-genes, samples of which are in the DNA (or somewhere else in the body or bodies). The ''battle'' between two meta-genes inside the same individual forms the quantum dynamics of psi-function at the level of the whole population. Possible results of the meta-genes battle within a single individual: one of the meta-genes goes into a latent state, and the second becomes dominant, or both become dominant and turn into different bodies or remain in one body. This hidden battle - ''private quantum of evolution'' is being protected from prying eyes and has a sharp character: this is ''peak explosion''. We thus can not apply Schrodinger equation directly to one individual, but after averaging over the population quantum language becomes the best way to express its evolution.

The number  $ P (g) $ of individuals with meta-gene $ g $ satisfies
Born rule $P(g)=|\Psi (g)|^2$. It does not characterize the specific individuals belonging to the population. Characteristics of individuals is connected with the phase $ \phi $ of wavefunction: $\Psi (g) = \rho (g) exp(i\phi (g))$. It is known from the hydrodynamic representation of quantum mechanics  (see. \cite{Bo}), that the phase $\phi$ is connected with the impulse $p$ of quantum swarm by the relation $p=\nabla\phi$,
e.g., the impulse is the gradient of the phase along spatial coordinates.  Impulse  $ p (r) $ is defined as the sum of impulses of all samples in a small cell containing a point $ r $.

It is necessary to define what is ''the impulse of the given meta-gene''.  This impulse is its aspiration to change, that is the ability of meta-gene for the transition $ g_i \ar g_j $ per unit time. However, such capacity obviously depends not only on the meta-gene itself, but from all of its carriers, i.e. those individuals who possess this meta-gen, from their states, placement in space, contact with the environment, which may include other organisms. Each meta-gene obtains its certain aspiration to change by summing of the results of all ''privat quanta of evolution''.

{\bf Impulse of a separate carrier of meta-gene is a measure of its aspiration to change in a certain direction.}

We can assume that the interaction of organisms occurs only when they are at one spatial point $x$. We add to each meta-gene $ G $ macroscopic spatial position $ x $ of beings that carry this meta-gene. We call a pair $(G,x)$ a local meta-gene with coordinates $G,x$. We can talk about the interaction of local meta-genes, if handled with the coordinates $ G $ in the same way as with ordinary spatial coordinates $ x $.

Thus, the module of wave function is determined by the distribution of population by local meta-genes, and the phase - by the individual organisms and the specific conditions in which they are.

Let us specify the foregoing as the following theses.

\begin{itemize}
\item Any state of the population is the psi function on the set of local
meta-genes, square of its modulus $ | \Psi (G, x) |^2 $ on the local meta-gene
$(G, x) $  is proportional to the total number of organisms with such meta-gene, averaged over time interval $ \tau $, and the phase gradient is equal to the sum of aspirations of these organisms to change their meta-gene.
\item If we introduce the evolutionary timescale $t=t_{phys}/T$, where $t_{phys}$ - is the usual time, in seconds, and record all values of $\Psi$ on all local meta-genes $(G,x)$ as a column $|\Psi\rangle$, then the evolution of this vector-valued function in time will
satisfy Schrodinger equation:
\end{itemize}
\begin{equation}
i|\dot\Psi (t)\rangle=H(t)|\Psi (t)\rangle
\end{equation}
where $H$ - is Hermitian operator slowly time dependent.

Let the set of classical states of meta-genes be organized 
as the space $ R^m $ with orthonormal basis $ e_1, e_2,
\ldots, e_m $, and for any classical trajectory in this space we have defined what it means
\nn

{\it aspiration to change the local meta-gene $ (G, x) \in R^m $ in
direction $ e_j $ equals real number
$\zeta_j$,}
\nn

so that all possible directions of drift $ e_1, e_2, \ldots, e_m $ not
depend on each other. Then the vector with components $ \zeta_j $ will be
impulse of this local meta-gene.

Let $ m_j $ denote the measure of resistance to movement along $ e_j $ from the side of meta-gene.

Quantum mechanics requires that the operator $ H $ has the form
\begin{equation}
H=\sum\limits_j-\frac{\Delta_j}{2m_j}+V(G,x)
\label{ham}
\end{equation}
where the operator $V$ acts on psi-function as a simple multiplication.
 In particular, $ V $ includes mutagenic factor.
An important condition is that if the population is not experiencing
any deterrent effect of the environment (i.e., occurs only under the influence of internal factors), then $ V = 0 $.

The general solution of Schrodinger equation $ \Psi (t) = exp (-iHt) \Psi (0) $ allows
 in principle, to simulate the evolutionary process in biology at
 population level and at large times $ t> T $. The price of such possibility - the need to consider complex meta-gene, which can be distributed on different bodies.

Application of the formula (\ref{ham}) requires a large density
of possible states of meta-genes, making it difficult to use. In addition, the practical value has simulation on short time intervals comparable with the life of the individual $t\approx \tau <<T$. In this case it is convenient to use quantum hydrodynamics of Bohm and simulate explosions of ''peaks'' in which the speed flight of individual samples will be $(\delta x_j)^{-3}\nabla_j\rho$ where $\delta x_j$ is the grain of spatial resolution in the space of meta-genes along the coordinate $j$  (\cite{KO}). 
The explosion will be the sharper, the more precisely we localize meta-gene.

The explosive nature of the evolution is a consequence of Bohr- Heisenberg uncertainty principle, which, in turn, follows from the more general rule: {\it any sample of a swarm has complete ''freedom of will'' when choosing path independent from other samples.} The stronger we
 localize the real system, the greater will its behavior differ from the classical one. This is reminiscent of the behavior of people in the crowd when its density is high (e.g., in the subway at rush hour), the crowd behaves very predictable, and you are ''adrift'', almost without thinking about the route. But if you - on a desert island, even the meeting with a person can plunge into shock (remember Robinson Crusoe): you do not know what's on his mind.

Scattering of samples of multiparticle systems - quantum worlds in the explosion is the exact formalization of the ''meta-gene'' selfishness, it can be treated as an uncontrollable striving for expansion. We cannot replace this property by some kind of ''altruistic'' behavior, where the samples, on the contrary, sought to be, so to speak, closer to each other. Altruism arises, for example, if we include into consideration other pieces of the gene that is expand it to meta-gene. Semblance of altruism can occur if the amplitude is concentrated on a certain state, as occurs, for example, in Grover search algorithm. In such cases, the ''altruism'' is not a primary feature of the population, but it follows from a specially arranged control of its state, which is implemented as a dependency of $ V $ of the time.

Suppose that there is a factor leading to the death of meta-genes that
 got to the point with coordinates $(G_d,x_d)$. If the population was concentrated in the initial instant in the point $ (G_i, x_i) $, ''explosion'' of this peak quickly lead to its ''spreading'' over the neighborhood of
starting point, so that some samples reach the point of death $(G_d,x_d)$. There will be the same as what happens with quantum
particle, which has met infinitely high potential barrier: its
wave function will tend to zero when approaching the barrier (total decline of samples does not matter - it is offset by the renormalization). Individual samples caught in the vicinity of the point of death, will receive a major boost in the opposite direction, that form a series of peaks of swarm density on the approach to the point of death; namely ''explosions'' of these peaks will remain unchanged form of the wave function similar to $ sin^2 $ in the neighborhood of zero (see picture 1), if average population size over a large period  $ T $ of time. Of course, we do not see the peaks themselves, as we do not see losers who died in the struggle for existence.

This form of the wave function will exist until such time as under the action of the external environment the death of the individuals stops at this point, then the population begins to propagate here as well.

But if we slightly lower the potential barrier, making the point
$ (G_d, x_d) $ not deadly, some samples can
overcome the barrier. These samples will have a high impulse in the direction to that point. If there is the zone behind the barrier, favorable for survival, the population will accumulate there, and even all will go exactly there, in spite of the barrier height (height only affects the transition time). This is a ''bottleneck'' of the evolutionary process (see \cite{Ma}), where
further evolution becomes unpredictable; here the small denominator in (\ref {hydro}) plays a role.

The opposite is true. If the point $ (G_w, x_w) $ is particularly favorable for growth of the population, samples will automatically start to accumulate there, even if at first they were not there. Moreover, if evolution was purely classical, not quantum, the population could not find a favorable point so quickly. For example Grover search algorithm finds the minimum point in a time of the order square root of the time of classical direct search.  

\section{Interaction of quantum meta-genes in food
 chain of the type ''wolves'' +''deer''}

Consider the set of all possible meta-genes, consisting of two
parts, the first we call ''wolf'', the second "deer'' and let
classical states of ''deer'' and ''wolves'' belong to
the set $\{ 0,1\}$. For ''deer'' $ 0 $ means a healthy state of
 meta-gene, in which it is protected against predators (this - not only
 the state of its genome, but also its location), $ 1 $ -
 a condition which makes it a potential prey. For a ''wolf'' $ 0 $  means the state of meta-gene that allows him to eat ''deer'', $ 1 $ - unreadiness for hunt. The values of all complex meta-genes $00,\ 01,\ 01,\ 11$ are defined respectively.

At the individual level meta-gene modification does not necessarily
mean a change of the DNA, for example, a state where ''deer'' can not react quickly to danger - such condition is temporary. But with the generations such transitions can lead to a real change in the genotype of real deer and real wolves. 

We will encode the state of our system by numbers $ 0,1,2,3 $ 
respectively, considering binary strings as arithmetic notation of these numbers. The matrix element $h_{i,j}$ is the amplitude of transition from the state $j$ into the state $i$ and vice-versa (Hamiltonian must be self-conjugated).  For example, the amplitude $h_{01}$ corresponds to the transition of the safe state of ''deer'' to the unsafe and vice-versa when the ''wolf'' is ready to hunt.  

Meaning of the state $ 11 $: ''deer'' is present only as a food eaten by a''wolf''. This makes the process irreversible at the individual level that on the language of quantum mechanics means that our system is open. Physically, this means that measurements occur nonrigid, which transform the state $11 $ to the state $00$: soil is fertilized, the grass grow, the other deer eat it, etc.
 This is an open system, whose dynamics is described by the generalization of Schrodinger equation for open systems: equation of Kossovsky - Lindblad:
\begin{equation}
\dot\rho (t)=-i[H,rho]+\sum\limits_{n, m = 1}^{N^2 - 1} b_{n, m} (L_n \rho L_m^+-1/2 (\rho L_m^+L_n + L_m^+L_n \rho))
\end{equation}
Unknown function will no longer be psi-function but the so-called
 density matrix $\rho$, which for the isolated system with pure state (psi-function)  $\Psi_k$ has the form $|\Psi_k\rangle\langle\Psi_k |$, and for the open system $\sum\limits_kP_k\rho_k$ that is the sum of density matrices of pure states taken with the probabilities $P_k$, $b_{i,j}$ - intensity of interaction with the environment. Diagonal of density matrix is formed by probabilities of basic states (quantities of the respective groups in the population). Here $L_n$ designates operators of projector type. In our case there is the single operator $L_1=|1\rangle\langle 3|$, which transform the state $ 11 $ - ''eaten'' deer into the state $ 00 $ - "healthy'' equilibrium state of the population; $b_{1,1}$- is nonzero intensity of the transformation   $11\ar 00$, all other intensities $b_{ij}$ equal zero. 

We take the following values for matrix elements: $h_{01}=0.01,\ h_{02}=0.01,\ h_{03}=0.05,\ h_{12}=h_{1,3}=0,\ h_{2,3}=0.01,\ h_{00}=h_{11}=h_{22}=h_{33}=1.$ 
Then the numerical simulation of the dynamics by Lindblad-Kossovsky
give the graph shown in picture 4. 

\begin{figure}
\includegraphics[scale=0.4, 
bb=5 10 483 500]{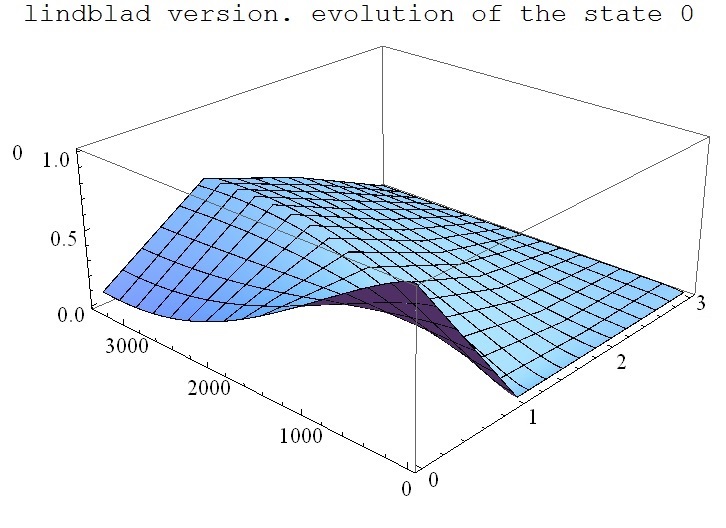}
 \caption{The graph shows the dynamics of the density of meta- genes. Numbers 0,1,2,3 correspond to initial states,
the simulation contains 3400 steps. 
The density of state 0 (''deer'' alive and
invulnerable,  ''wolves '' are fed) is stabilized at around 0.1, state 1 (''deer''
alive but vulnerable ''wolves'' are hungry) at 0.6 and state 2 
(''deer'' vulnerable, ''wolves'' fed) at 0.3}
\end{figure}

\section{Unitarity and singularity}

The evolution operator $U_t:\ {\cal H}\ar {\cal H}$ acts on the space of quantum states, and its unitarity means preserving the distances between these states. This is - a much deeper property than those formulated in terms of
separately taken states. For biology, this means that the evolution preserves the distance between any two possible states of the entire population. It is important that the evolution operator acts immediately on the whole bundle, consisting of closely lying possible paths of evolution. 

Precise knowledge of the wave function is an abstraction that has
direct sense only for such systems, which identical quantum
states we can prepare in great numbers in order to collect statistics. For living organisms one cannot even think about this. Therefore, the singularity (''peaks'') play in the biological evolution decisive role.

These ''peaks'' retain the shape of the stationary wave functions in the
model tasks of quantum mechanics, just like a forest growing on
hillside keeps it from collapsing.
Unitarity - a sign of good management of population. A good
 shepherd retains herd size.  Wolves do not eat his sheep, and the numbers are not decreasing and the sheep do not eat vegetables growing in the garden, and therefore there is no uncontrolled reproduction. 

But if the number is not saved (as in the example from the previous
section), it means that someone from the outside really controls. In this case, the meta-gene must be expanded to a level where it will once again be the unitary evolution. If the ''wolves'' have eaten all ''deer'', they will have to move to another food, for example, start eating fish. Meta gene will still consist of a pair of qubits, only semantics will be different. But at the global
level unitarity must always remain! 

\section{Conclusions}

Brief statement of our findings is as follows:

\begin{itemize}
\item Selfishness of biological genes has a quantum nature, and can be described as ''explosions'' of peaks of their carriers density.
\item Quantum nature of the gene is responsible for the biological evolutionary process in the following precise sense: the density of the wave function of the meta-gene corresponds to the number of its carriers, and the aspiration to change, depending on the specific creatures, corresponds to the phase of the wave function.
\item Unitary quantum operator describing biological evolution, requires a combination of several meta-genes in the same individual, and distribution of a meta-gene by the different bodies.
\end{itemize}

We have seen that quantum mechanics determines aspiration to expansion of the gene that is characterized as ''selfishness'' of its behavior. The uncertainty principle of Bohr -Heisenberg is the hidden spring of irrepressible aspiration to expansion, which shows all alive. We can hope that the quantum-mechanical representation of biological evolution one day will help us to find its right direction with respect to ourselves.

\section{Acknowledgements}

Howard Blum drew my attention to the possibility of a close connection between the collective behavior of quantum many-body systems and behavior of living beings. From his book \cite {Bl} I learned about Dawkins ideas. I am also grateful to my mother,  biologist Aida Ozhigova, discussions with which have greatly contributed to the growth of my knowledge of biology and the theory of evolution. 

The article is written with the support of the Russian Foundation for Basic Research, grant 12-01-00475-a.

\end{document}